# Hybrid Metal-Dielectric Plasmonic Zero Mode Waveguide for Enhanced Single Molecule Detection

*Xavier Zambrana-Puyalto,[1] Paolo Ponzellini,[1] Nicolò Maccaferri[2], Enrico Tessarolo,[3] Maria G. Pelizzo,[3] Weidong Zhang,[4] Grégory Barbillon,[5] Guowei Lu,[4] and Denis Garoli[1*]*

[1] Istituto Italiano di Tecnologia – Via Morego, 30, I-16163 Genova, Italy
[2] Physics and Materials Science Research Unit, Université de Luxembourg, L-1511 Luxembourg, Luxembourg
[3] CNR-IFN, Via Gradenigo 3, Padova (Italy)
[4] State Key Laboratory for Mesoscopic Physics & Collaborative Innovation Center of Quantum Matter, School of Physics, Peking University, Beijing 100871, China
[5] EPF—École d'Ingénieurs, 3 bis rue Lakanal, 92330 Sceaux, France
[*] Corresponding author's email: denis.garoli@iit.it

**Abstract**

We fabricated hybrid metal-dielectric nanoantennas and measured their optical response at three different wavelengths. The nanostructure is fabricated on a bilayer film formed by the sequential deposition of silicon and gold on a transparent substrate. The optical characterization is done via fluorescence measurements. We characterized the fluorescence enhancement, as well as the lifetime and detection volume reduction for each wavelength. We observe that the hybrid metal-dielectric nanoantennas behave as enhanced Zero Mode Waveguides in the near-infrared spectral region. Their detection volume is such that they can perform enhanced single-molecule detection at tens of μM. However, a wavelength blue-shift of 40 nm dramatically decreases the performance of the nanoantennas. We compared their behavior with that of a golden ZMW, and we verified that the dielectric silicon layer improves the design. We interpreted the experimental observations with the help of numerical simulations. In addition, the simulations showed that the field enhancement of the structure highly depends on the incoming beam: tightly focused beams yield lower field enhancements than plane-waves.

**Introduction**

Single molecule fluorescence is a broadly used technique in biophysical research, in particular in diagnostics and sequencing[1–4]. In order to reach the single molecule detection limit, a careful optimization of the signal-to-noise ratio is required. Several schemes for reducing the fluorescence background from out-of-focus probes have been investigated[5–11]. Two of the main techniques to reduce out-of-the-focus background are total internal reflection fluorescence and confocal microscopy. The main limitation for their use in single molecule spectroscopy is that nanomolar concentrations are needed[12–14]. One of the most powerful technologies for single molecule fluorescence spectroscopy at concentration close to biological processes (micro and millimolar) is based on the use of Zero Mode Waveguides (ZMWs)[14,15]. Ever since the pioneering work of Levene *et al.*[16], ZMW have received a lot of attention[17,18]. Their ability to reduce the optical detection volume by several orders of magnitude (from $10^{-15}$ liter with a standard confocal microscope to $10^{-21}$ liter) allowed for carrying single molecule studies on chips made of hundreds of nanoholes[16,19]. Most of the used ZMWs are prepared as circular holes in thin aluminum film[20]. Yet some alternative shapes and materials have been proposed to improve the detection efficiency of these structures[21,22]. One of the most successful alternative designs has combined the field enhancement of optical nanoantennas and the fluorescence screening of plasmonic nanoapertures[23,24]. As for the materials, it has been seen that the antennas fabricated on Ag or Au films yield a better electromagnetic (EM) field confinement, which can lead to a significant enhancement in fluorescence emission with respect to the one achieved in Al ZMWs[25–27]. Unfortunately, the use of plasmonic antennas as ZMWs has some important drawbacks: (i) a circular or rectangular nanoantenna in Au or Ag can typically confine the field within the cavity but is not so efficient in the confinement at the bottom interface between the nanoholes and the transparent substrate; (ii) the enhanced field confined close to the metal interface can introduce additional quenching effects in the dye emission. In order to circumvent these limitations, alternative designs such as bilayered nanoantennas have been proposed. For instance, Zhao *et al.*[26] firstly proposed a bilayer nanoantenna on Au-Al substrate. The design was recently experimentally verified by some of us, demonstrating a significant fluorescence enhancement with respect to a standard Al ZMWs[27]. Lu *et al.* also proposed a hybrid metal-dielectric antenna where a Si/Au bilayer is used to achieve better field distribution and higher antenna quantum yield with respect to the same pure-gold nano-aperture[28]. Inspired by this design, we have fabricated ZMW nanostructures on a bilayer of Si-Au on a glass substrate. We fabricated both circular nanoholes and rectangular nanoslots verifying that the latter perform better (in terms of fluorescence enhancement) as expected from some of our previous works[27]. Hence, we focus our investigation on rectangular nanoslots excited at three different wavelengths: 587, 633 and 676 nm. For each wavelength, we performed a reference experiment on a cover glass. We measured the fluorescence enhancement, the reduction of the detection volume, as well as the lifetime reduction with respect to the reference confocal measurement. We used Alexa610, Alexa647 and Alexa680 at each of the respective wavelengths. We observed that a blue-shift of 40 nm (from 676 to 633 nm) is enough to completely change the behavior of the ZMW. We performed numerical simulations in order to understand this effect. We have seen that the effect is due to a change in the field structure, which goes from being confined at the bottom of the structure to being almost spread across the whole volume of the nanoslot. We compared the behavior of the hybrid metal-dielectric nanoslot with a gold nanoslot of the same dimensions. We observed that the hybrid metal-dielectric nanoslot outperform the gold nanoslot in the near-infrared. As put forward by Lu *et al.*[28], we mainly attributed these differences to the fact that Si absorbs less than Au in the near-infrared, and consequently there is less quenching. Last but not least, we showed that the intensity enhancement produced by these types of structures heavily depends on the incoming beam used to produce it. We showed that tightly focused Gaussian beams yield lower intensity enhancements and intuitively explain why.

**Results and Discussion**

The experimental verification of the design proposed by Lu et al. (Fig. 1) started with the fabrication of the nanostructures. Following the procedures reported in methods, we fabricated rectangular nanoslots (35x100nm wide at the nanoslot-glass interface). We fabricated these nanoslots both a Si-Au bilayer. The total thickness is 100nm. The preparation of 35 nm wide structures into a 100 nm thick metallic film can be performed following different strategies such as electron beam lithography and lift-off, or direct milling of the structure by means of focused ion beam (FIB). The first method was discarded because it presents some difficulties during the lift-off process, due to the high transverse aspect-ratio and to the small dimensions of the structures. Instead, the FIB allows for a rapid preparation of the nanoslots by directly milling into the Si-Au film. An intrinsic drawback of FIB milling is the unavoidable flared shape of the fabricated slots. The cross section in Fig. (1c) shows this phenomenon for the Si-Au case. A second issue related to the FIB fabrication regards the milling depth. The milling time needed to be tuned in order not to dig into the glass substrate below the metallic layer (methods).

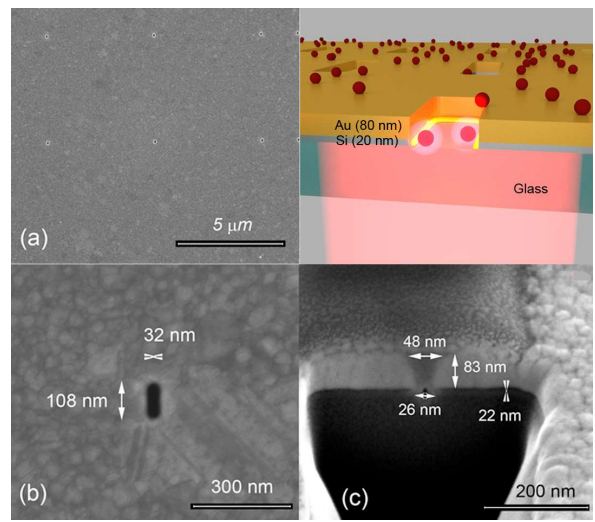

**Figure 1**: SEM micrographs of the investigated structures. (a) Top view of the array; (b) top view of a single antenna; (c) cross-section in order to illustrate the geometry and the thin layers. Right-top-panel schematic illustration of the structure.

Next, we set out to experimentally test the fabricated nanoantennas for three different wavelengths. The aim of these measurements is i) to demonstrate that the hybrid metal-dielectric ZMWs proposed by G. Lu *et al.*[28] are a good choice of ZMW for the near-infrared, and ii) to see how its behavior changes for visible wavelengths. For that, we used the same optical set-up used in some of our previous works[27,29]. The illumination is obtained from a supercontinuum laser, which is coupled to a single mode fiber. Unfortunately, our current set-up cannot perform fluorescence correlation spectroscopy (FCS) or lifetime measurements for laser excitations above 680 nm. Hence, we chose the following central wavelengths for the experiments: 587, 633 and 676 nm. The power is maintained at 25 µW for the three wavelengths. The beam is focused onto the sample with an oil-immersion microscope objective (MO) with a numerical aperture (NA) of 1.3. As depicted in Fig. 1, the incoming light hits the sample through the dielectric part. Three different dyes are used for the respective wavelengths: Alexa610, Alexa647 and Alexa680. The fluorescence from the dyes is collected with the same MO. The backscattering from the sample is removed using optical filters (see Methods). Finally, the fluorescence signal is split into two paths by a 50/50 beam splitter, and each of the respective signals is detected by an APD (see Methods).

Once the set-up has been aligned, we carry out an FCS and a lifetime measurement on top of a cover glass. We name this configuration the confocal case. These measurements give us the volume of detection as well as the count rate per molecule (CRM) and the lifetime of the dyes in this confocal case. We used these measurements as a reference for each wavelength. Afterwards, we performed the same measurements with the ZMWs and compared the results. The ratio between the values of CRM, volume and lifetime yield the following three figures of merit: Fluorescence enhancement (FE), volume reduction (VR) and lifetime reduction (LR). We expressed the ratio of the detection volumes as well as the lifetimes as reduction (instead of enhancement) as we would like a ZMW to reduce the detection volume and the lifetime emission. While Figure 2 illustrates some examples of experimental data, in Table 1, we give the three figures of merit for the fabricated nanoantennas at the three wavelengths of study. The details about the FCS and lifetime measurements are given in Methods.

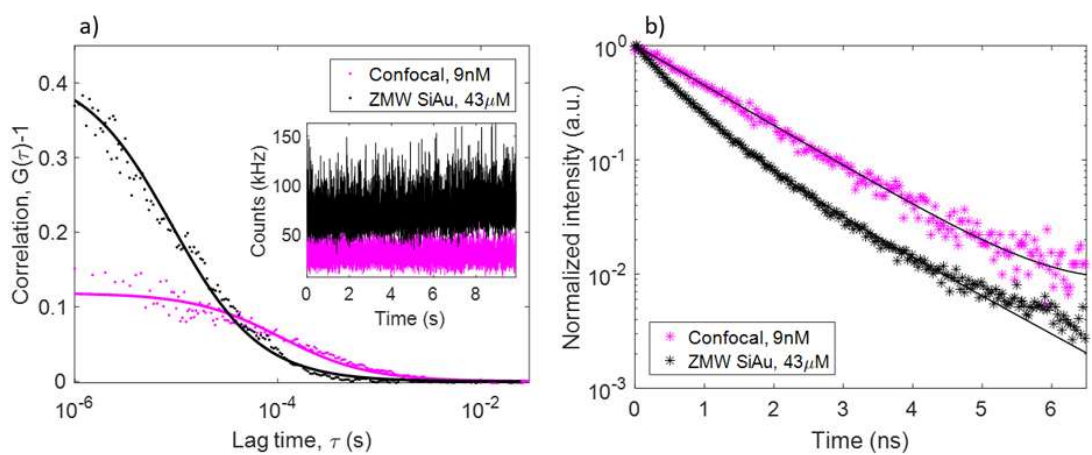

**Figure 2**: a) FCS measurements. The correlations have been obtained with the fluorescence time traces given as an inset. b) Normalized decay traces. The confocal measurements (in magenta) have been obtained with 9 nM of Alexa680. The measurements using the hybrid SiAu ZMW (in black) have been obtained with 43 μM of Alexa680. The laser excitation is at 676 nm, and the fluorescence is measured in between 695-725 nm.

|    | 587nm        | 633nm        | 676nm         |
|----|--------------|--------------|---------------|
| FE | 1.03 ± 0.14  | 1.34 ± 0.19  | 10.5 ± 1.2    |
| VR | 90 ± 40      | 560 ± 170    | 12800 ± 5500  |
| LR | 1.02 ± 0.05  | 1.95 ± 0.14  | 2.45 ± 0.01   |

**Table 1**: Summary of the three figures of merit for the hybrid metal-dielectric nanoantennas.

ZMWs are especially defined by their ability to concentrate the EM field in very small volumes at the bottom of structure. This is their main feature, as they allow for doing single molecule spectroscopy at µM concentrations. The experimental measurements summarized in Table 1 confirm the fact that the hybrid metal-dielectric nanoantennas behave as ZMWs in the near-infrared. We observed that despite using a concentration of 50 µg/mL (~43µM) of Alexa680, we measured an average number of molecules $\langle N \rangle \approx 1$ under a 676 nm beam excitation. As a result, we obtained a VR of the order of $10^4$. Furthermore, we observe that the hybrid metal-dielectric give a FE of 10.5, which is one magnitude order larger than the typical Al ZMWs[27]. As for the lifetime reduction, a factor of 2.5 is in accordance with similar ZMWs made in Al or Au[22]. As expected, we see that the performance of the nanoslots as ZMWs deteriorates as we blue-shift the wavelength. When the excitation wavelength is changed to 633 nm, we observe that the detection volume goes down by a factor of 20. Moreover, the FE given by the structure almost vanishes. Still, looking at the three figures of merit, we can conclude that the use of the nanoantenna to probe single molecules at 633 nm is a clear improvement with respect to the confocal case. Nevertheless, the last claim cannot be made for the nanoantennas working at 587 nm. As shown in Table 1, using the hybrid metal-dielectric nanoantennas at 587 nm do not yield any lifetime reduction nor any fluorescence enhancement. Besides, the volume reduction is only in between one or two magnitude orders, and the measurement has a large uncertainty. Finally, note that the fluorescence measurements are not done at 587, 633 and 676 nm, but rather in between 600-680 nm, 660-700 nm and 695-725 nm.

Next, we fabricated a set of nanoslots on a gold layer (see Methods). The nanoslots have the same shape as the hybrid metal-dielectric nanoslots. The only difference between this design and the one reported in Fig. 1 is the lack of the Si layer. (SEM micrographs of these structures are given in the SI). As it was theoretically shown by G. Lu *et al.*[28], the main role of the Si layer is to prevent the dyes from quenching, as Si absorbs much less than gold in the infrared. In Table 2, we present the results of the FCS and lifetime measurements carried out with the golden nanoslots. The measurements have been done for the same wavelengths, and using the same filters and dyes. A reference measurement in the confocal configuration has also been done.

|    | 587nm        | 633nm         | 676nm         |
|----|--------------|---------------|---------------|
| FE | 1.3 ± 0.5    | 1.1 ± 0.2     | 2.8 ± 0.3     |
| VR | 5 ± 2        | 1500 ± 500    | 6000 ± 600    |
| LR | 1.02 ± 0.08  | 1.91 ± 0.07   | 2.10 ± 0.08   |

**Table 2**: Summary of the three figures of merit for the golden nanoantennas.

We observed that the qualitative spectral behavior of the gold nanoantennas is the same as the hybrid ones. That is, the nanoantennas behave as ZMWs at 676 nm, and this behavior is partially lost when the excitation wavelength is blue-shifted to 633 nm. Then, at 587 nm the ZMW behavior is completely lost.

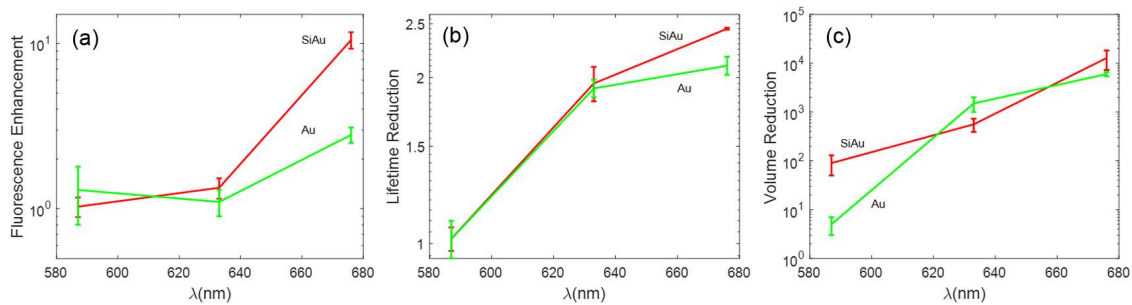

**Figure 3**: Figures of merit for the hybrid metal-dielectric (SiAu, in red) and the golden (Au, in green) nanoantenna. The three figures of merit are plotted in a) FE, b) LR, and c) VR.

In order to quantitatively compare the two nanoantennas, we plot the three figures of merit for the two designs in Figure 3. Here, we clearly see that the Si layer makes a big difference in the infrared. The FE is almost an order of magnitude higher, and the VR reduction is double. In contrast, the LR is just a 15% higher. We believe that the differences in the three figures of merit would have been greater if we had done the experiment for longer wavelengths, as the absorption of Si decreases with the wavelength. Unfortunately, we have not been able to experimentally prove it as our optical set-up does not allow us to work for longer wavelengths than 680 nm.

In order to better understand the experimental results, we simulated the optical behavior of the Si-Au nanoslots by means of FDTD (Lumerical - see Methods). The simulated structure can be seen in Fig. 4, where the illumination travels upwards, and the solution containing the dyes is on the gold side. We show the near-field intensity distribution upon a tightly Gaussian beam illumination (see Methods).

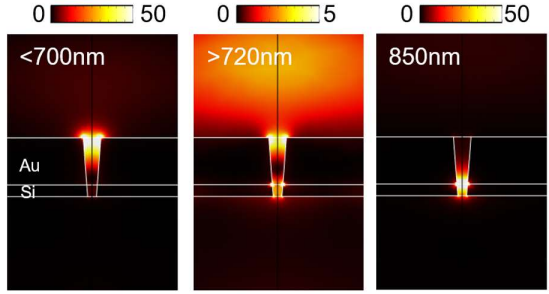

**Figure 4**: Electric Field Intensity ($|E^2|$) distribution expected for spectral range below 700 nm; above 720 nm and at 850 nm respectively.

We observe that, depending on the wavelength, the illumination creates a significantly different intensity distribution in the nanoslot. More specifically, we observe that for wavelengths below 700nm, the intensity distribution is mainly concentrated at the top of the structure, *i.e.* on top of the gold layer. In contrast, for 720 and especially for 850 nm, the intensity distribution resembles that of a typical ZMW where the field is pressed down to the bottom of the structure. Clearly, the performance of the structure as ZMW will be better for the longer wavelengths.

Notice that all the simulations have been done using a tightly focused Gaussian beam. The reason is obvious: the experimental excitation of the nanoantennas is done with an oil-immersion objective (NA=1.3), which creates a tightly focused Gaussian beam. Still, one could wonder if the choice of this illumination, instead of a broadly used plane-wave, makes any difference to compute intensity enhancements (IE). Here, we show that it does make a significant difference, quantitatively speaking. There have been many different works showing that the scattering of a nanostructure can be radically changed by tailoring the illumination[30–36], but none of these works has quantified the differences in near-field enhancement. In Figure 5, we compare the IE spectra of the hybrid metal-dielectric nanoslots excited with a tightly focused beam and with a plane-wave. We can see that their spectral properties are qualitatively the same ones. However, the IE produced by the plane-wave is approximately one magnitude order greater than that produced by the tightly focused beam. The exact factor between the IE with a plane-wave and a tightly focused Gaussian beam depends on the nanostructure in consideration. Yet, here we intuitively explain why the IE produced by a tightly focused Gaussian beam will generally be lower than that produced by a plane-wave. The underlying reason is that the local intensity of a tightly focused beam at its focal plane is much greater than that of a plane-wave in the same focal plane. The IE is computed either in one point, surface or volume, which is representative of the nanostructure in consideration. Then, the IE is computed in the chosen region as a ratio between the near-field intensity produced by the illumination, and the illumination itself (when there is no structure). And the big difference between the two cases is that, as mentioned before, the local intensity of a tightly focused beam is significantly greater than that of a plane-wave. This is because the intensity of a plane-wave uniformly spreads all over space, whereas the intensity of a tightly focused beam is confined in a small region.

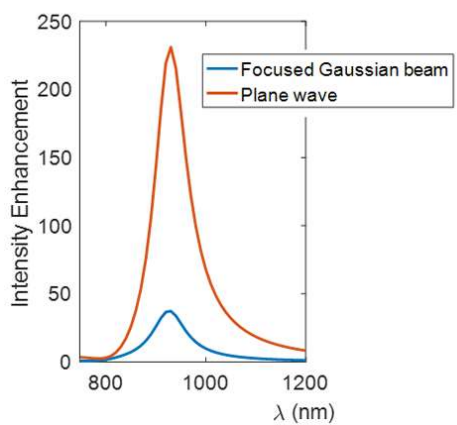

**Figure 5:** Comparison between the intensity enhancement induced by a tightly focused gaussian beam and a plane-wave.

**Conclusions**

In summary, we fabricated hybrid metal-dielectric nanoantennas that perform as optimized ZMW in the near-infrared. We measured their performance at three different wavelengths (587, 633, and 676 nm) and have seen that the ZMW behavior can be lost when the wavelength is shifted by 40 nm. We carried out some simulations that have showed us that this change in behavior is due to the fact that the field confinement changes from being at the bottom of the nanoslot, to being at the top of the gold surface. In addition, we have seen that the illumination plays a crucial role in order to quantify the intensity enhancement. In particular, we showed that a tightly focused Gaussian beams induced an intensity enhancement that can be one magnitude order below that created by a plane-wave. Last but not least, we fabricated nanoantennas on a gold film with the same design. We have experimentally compared the performance of the two families of nanoantennas. We observed that the metal-dielectric nanoantennas clearly outperform the ones made of gold in the infrared.

**Methods**

Sample fabrication

150 μm thick glass slides have been used as the substrate. Firstly, the substrates have been thoroughly cleaned in order to obtain a smooth metallic surface with the subsequent metal deposition. The substrates have been washed in a piranha solution (3:2 $H_2SO_4$:$H_2O_2$), thoroughly rinsed in deionized water, and subsequently sonicated in acetone and in isopropanol. The substrates have been dried by means of a nitrogen gun. The metal layers, then have been deposited by means of electron-beam evaporation in high-vacuum. The use of a thin layer of titanium is typically required in order to avoid the delamination of the metallic layers from glass, but in this experiment we observed that silicon acts as a good adhesion layer for both gold and glass. Hence, for the hybrid Si-Au nanoslots, a 20 nm silicon layer has been evaporated directly on the glass substrate, at a deposition rate of 1 Å/s, and a 80 nm layer of gold has been evaporated above the silicon layer, at the rate of 0.3 Å/s, without breaking the vacuum in the vacuum chamber, between the two depositions.

The bilayer has been drilled by means of a gallium FIB at an accelerating bias of 30 kV, with a dwell time of 1 μs, and with the ionic current set to a low value (24 pA) in order to obtain a tiny beam and hence narrow structures. Due to the high aspect ratio of our structures, the actual inner shape of the nanoslots is not visible by means of a top view SEM image. Only the upper, external shape of the nanoslots, at the gold-water interface, results clearly visible and measurable. For evaluating the milling parameters, both in the vertical and horizontal planes, several cross-sections have been realized on trial nanoslots. First of all, this approach has allowed us to calibrate the milling time, in order to reach the glass substrate below the metallic layer, without digging into the glass. In the second place, it has allowed us to measure the resulting slope of the nanoslot walls. This way we could calibrate the design of the nanoslots within the FIB software in order to obtain the desired rectangular shape with the optimal simulated dimensions (35x100nm), at the nanoslot bottom interface.

The gold nanoslots have been fabricated with the same procedure, with the obvious exception of the material deposition. Even in this case the deposition was performed in an electric beam evaporator, in high vacuum. But here a 3nm thick titanium layer has been used as the linker. Titanium has been deposited at 0.3 Å/s. Above the titanium layer, a 100 nm gold layer has been deposited, still at 0.3 Å/s, without breaking the vacuum, between the two depositions.

Once fabricated, the samples have been stored in a nitrogen atmosphere to avoid deterioration. Before the fluorescence measurements, a 90 seconds oxygen plasma (100% $O_2$; 100 W) was performed on the sample. The plasma treatment is fundamental to make the surface hydrophilic, and hence to ensure that water enters the nanoantennas.

Optical set-up

The experiments are performed on an inverted microscope with an oil-immersion (NA=1.3) microscope objective. The light beam, which enters the microscope through its rear port, is obtained with a supercontinuum laser, which has been previously coupled to a single mode fiber. Three different central wavelengths have been used: 587, 633, and 676 nm. The linewidths of each case are 15, 17, 20 nm, respectively. All the experiments have been performed at a power of 25 µW. The samples that we used for the experiment are held on a sample holder, which is attached to a micro and a nanopositioner. The nanopositioner is used to center the sample with respect to the incident beam. The nanopositioner is also used to place the sample at the z-plane where the fluorescence counts are maximized. All the samples are cleaned with an oxygen plasma at 100 W for 2 min before being used. For each of the wavelengths, a different dye of the Alexa family is used: Alexa610, Alexa647 and Alexa680. For the three cases, a 30µL droplet of the corresponding solution of dyes is used. The concentration is always kept at 50 µg/mL for the experiments with the ZMWs. The confocal experiments have been done with two different concentrations, 10 ng/mL and 100 ng/mL. Three filters are used to separate the fluorescence of the dyes from the light reflected off the sample. The filters are a dichroic mirror (DM), and a combination of either two bandpass filters (BP) or a longpass filter (LP) and a BP. For Alexa610, the filters were: DM at 594 nm, a LP at 633 nm, and a BP at 641/75 nm. For Alexa647: DM notch at 633 nm, and two BPs at 698/70 nm and at 680/42 nm. For Alexa680: DM at 685 nm, a LP at 685 nm and a BP at 711/25 nm. After the last BP filter, a 50/50 beam splitter is used to split the fluorescence light signal into two different channels. Then, two 50 mm lenses are used to focus the fluorescence signal onto two equivalent avalanche photodiodes (APD). The APDs are re-aligned with XY translation stages for each different wavelength. The signal of the APDs is recorded with a time-correlated single photon counting module.

FCS measurements

The fluorescence trace duration is typically 1 minute. The binning time has been maintained between 10 and 200 ns, depending on the case. After the detection of the fluorescence trace, three different calculations are carried out to be able to retrieve the count rate per molecule (CRM). First, the signals measured by the two APDs are correlated. A 2D translational diffusion model is used to fit the correlation function and obtain the average number of molecules $\langle N \rangle$[37]. This is a valid approximation of a 3D diffusion model when of the axial size of the observation volume is much larger than its lateral[38]. Second, the average of the fluorescence trace $\langle F \rangle$ and its standard deviation $\delta F$ are calculated. Finally, the CRM is obtained as $CRM = \langle F \rangle / \langle N \rangle \pm \delta F / \langle N \rangle$. This operation is performed several times, and the final CRM is obtained as an average of the different CRM measurements for each slot.

Lifetime measurments

The APDs as well as the laser are connected to a time-correlated single-photon counting module in time-resolved mode. Making use of a home-built code, we build a histogram of 300 bins, each of them having a temporal width of 30 ps. A certain time delay is applied to the laser channel, so that the histogram is monotonously decreasing. The histogram measurement is carried out for 150 s. A single exponential $Te^{-t/\tau}$ is used to fit the confocal histograms, whereas biexponential function of the kind $Ae^{-t/\tau_A} + Be^{-t/\tau_B}$ is used to fit the decay curve for the ZMWs. Typically, the $Ae^{-t/\tau_A}$ term contains information about the fluorescence, whereas $Be^{-t/\tau_B}$ is mainly noise. We checked that the contribution of $Ae^{-t/\tau_A}$ is at least one magnitude order of larger than that of $Be^{-t/\tau_B}$. Each measurement is repeated several times, and the result is given as the average plus an uncertainty given by the standard deviation.

**Author Contributions**
X.Z.-P. performed the optical measurements. P.P., E.T. and M.G.P. fabricated the samples. N.M., W.Z., G.B. and G.L. designed and simulated the structures. X.Z.-P. and D.G. wrote the manuscript. D.G. coordinated the work.

**Competing financial interests:**
The authors declare no competing financial and non-financial interests.